\documentclass[aps,pra,prabib,twocolumn,showpacs,nofootinbib]{revtex4}
\usepackage{graphicx} \usepackage{amsmath} \usepackage{amssymb}
\usepackage{amsfonts} \usepackage{bm} \usepackage{mathrsfs}

\begin{document}

\newcommand{\be}{\begin{equation}} \newcommand{\ee}{\end{equation}}
\newcommand{\bea}{\begin{eqnarray}}\newcommand{\eea}{\end{eqnarray}}


\title{Dirac spinors in solenoidal field and self adjoint extensions of its
Hamiltonian}

\author{Pulak Ranjan Giri} \email{pulakranjan.giri@saha.ac.in}

\affiliation{Theory Division, Saha Institute of Nuclear Physics, 1/AF
Bidhannagar, Calcutta 700064, India}

\begin{abstract}
We discuss  Dirac equation (DE) and its solution in presence of
solenoid (infinitely long) field in (3+1) dimensions.  Starting with
a very restricted domain for the Hamiltonian,  we show that a
1-parameter family of self adjoint extensions (SAE) are necessary to
make sure the correct evolution of the Dirac spinors.  Within the
extended domain  bound state (BS) and scattering state (SS)
solutions are obtained. We argue that the existence of bound state
in such system is basically due the breaking of classical scaling
symmetry by the quantization procedure. A remarkable effect of the
scaling anomaly is that it puts an open bound on both sides of the
Dirac sea, i.e., $E\in(-M,M)$ for $\nu^2[0,1)$! We also study the
issue of relationship between scattering state and bound state in
the region $\nu^2 \in[0,1)$ and recovered the bound state solution
and eigenvalue from the scattering state solution.
\end{abstract}



\date{\today}

\maketitle

\section{Introduction}

The problem of a Dirac particle in magnetic fields background has been
studied extensively in literature \cite{uniform,ger,park,fal}. The
solution in a uniform field  background \cite{uniform} is known
for a long time.
For a solenoidal background, the Dirac equation has been solved in
(2+1) dimensions \cite{ger,park,fal}. It shows that a self-adjoint
extension of the Dirac Hamiltonian is necessary to obtain the
solutions.  In (3+1) dimensions, self-adjoint extension of the Dirac
Hamiltonian has also been studied \cite{avo}, but it involves a
$\delta$-sphere potential rather than solenoid interaction. In
Ref.~\cite{gitman}, self-adjoint extension of Dirac Hamiltonian in
solenoid field and an uniform magnetic field $B$ has be studied in
both (2+1) and (3+1) dimensions elegantly.

The consequences of considering self-adjoint extensions for the
problem of Dirac particle in solenoidal field is recognized in Ref.
\cite{hagen}. Because SAE \cite{feher,tsutsui,veigy}
is a method which allows to consistently
build the all possible boundary conditions under which the
Hamiltonian is self-adjoint.  It is therefore expected that \cite{ger} the
results would be function of the boundary conditions, characterized by
parameters.

In this paper, we study the solution of the Dirac equation
in only infinitely long
solenoid field in $(3+1)$ dimensions. Because, the magnetic field of
an infinitely long solenoid is immensely important for various
reasons. For example, it is used to study the Aharanov-Bohm
\cite{bohm} effect. Similarity with this field also helps one to
understand the physics of cosmic strings \cite{string}. Although, in
principle problem in Ref.~\cite{gitman} should reduce to the problem
involving only solenoid field by making the uniform magnetic field
$B=0$. But it is a nontrivial task. So it is important to study
Dirac free Hamiltonian in only solenoid field background in (3+1)
dimensions and solve it with the most  generalized boundary
conditions, such that the Hamiltonian is self-adjoint. Some comments
about the difference between  method of \cite{gitman} and ours will
be made in Sec. VII.

In order to get the most generalized boundary condition for our
problem, we start with a very restricted domain. Obviously the
Hamiltonian is not self-adjoint in that domain. We then go for a
self-adjoint extensions of the Hamiltonian by using the von
Neumann's method \cite{reed}. We  map the problem from the Hilbert
space $L_2(dr d\phi dz; \mathcal C^4)$  to $L_2(dr d\phi dz;
\mathcal C)$ keeping in mind that we seek self-adjointness of the
whole $(4\times 4)$ Dirac Hamiltonian in solenoid field background.
In the Hilbert  space  $L_2(dr d\phi dz; \mathcal C)$  we get a well
known one dimensional Schr\"{o}dinger eigenvalue problem of inverse
square interaction. Inverse square interaction is known to be
classically scale invariant. But, quantization of this system shows
that the classical scale symmetry is broken for some values of the
self-adjoint extension parameter and for a very limited range
$-1/4\leq g<3/4$ of the coupling constant $g$ of the inverse square
interaction. Breaking of this scaling symmetry by quantization is
known as scaling anomaly. The indication of scaling anomaly in the
system is manifested through the formation of bound state, which is
supposed to be absent if scaling symmetry is present even after
quantization. In case of Dirac particle in solenoid field, this
scaling anomaly is also responsible for the formation of bound
state. It is also possible to comment on the bounds (upper and lower
bounds) of the bound state energy of the Dirac particle  on solenoid
filed background. By taking appropriate limit, it is even possible
to find the upper and lower bounds of the energy of the free Dirac
particle.

The paper is organized as follows: In Sec. II, we state the problem
of Dirac equation  (DE) in solenoid field background. In Sec. III,
we state the usual symmetric boundary condition using regularity and
square integrability argument and we perform the required self
adjoint extension. In Sec. IV,  we discuss the bound states of the
radial Hamiltonian and we get the bound state condition. In Sec. V,
we discuss the scattering states of the radial Hamiltonian and get
the corresponding condition for it. In Sec. VI, we discuss the
scaling anomaly present in the system.  We discuss in Sec. VII.
\section{Solutions of the DE in solenoid field}

We consider a Dirac particle of mass $M$ and charge $eQ$ in solenoid
field background. Due to the cylindrical symmetry of the problem it
is better to use the cylindrical polar coordinate system, where the
spatial coordinates are denoted by $r, \phi, z$.
In any orthogonal
coordinate system, the Dirac equation for a particle with charge
$eQ$ and mass $M$ can be written as Ref.~ \cite{gre}

\begin{eqnarray}
 \left[ \gamma^\mu \left( i\hat D_\mu - eQA_\mu  \right) - M~~~~~~~~~~~~~~~~~~
 \right.\nonumber\\ \left.+ i\sum_i \gamma_i\left[\frac{1}{2}\hat D_i
 \log\left(h_1h_2h_3/h_i\right) \right]\right]\Psi_r\ = 0,
\label{Dirac}
\end{eqnarray}

where $h_i's$ are scale factors of the corresponding coordinate
satem. In our case since we are using cylindrical coordinate system,
$h_1 =1, h_2 = r, h_3 = 1$. the derivative $\hat D_\mu =
(h_\mu)^{-1}\partial_\mu$, where no summation over $\mu$ is implied
here and $A_\mu$ is the infinitely long solenoid vector potential,
$\vec A = \frac{\alpha}{r}\hat\phi$. One can make a conformal
transformation \cite{gre}, which reduces  (\ref{Dirac}) to simpler
form with zero spin connection as \cite{gre},
\begin{eqnarray}
 \left[ \gamma^\mu \left( i\hat D_\mu - eQA_\mu  \right) - M \right]\Psi(t, r,
 \phi, z) = 0,
\label{dirac1}
\end{eqnarray}
where the relation between $\Psi_r$ and  $\Psi$  is $\Psi_r =
(1/\sqrt r)\exp(-\frac{1}{2}i\phi\Sigma_3)\Psi.$ $\Sigma_3$ is the
generator for rotation along $z$-axis \cite{gre}. The normalization
condition of the new wave-function $\Psi$  is
\begin{eqnarray}
\int dr d\phi dz \; \Psi^\dagger(t, r,\phi,z) \Psi(t, r,\phi,z) = 1.
\label{norm}
\end{eqnarray}
To solve the Dirac equation (\ref{dirac1})
we take the trial solution of  the Dirac equation of the form
\begin{eqnarray}
\Psi(t, r, \phi, z) = e^{-iEt}e^{-im\phi}e^{-ip_zz} \left( \begin{array}{c}
\phi(r) \\ \chi(r) \end{array} \right) \,,
\label{trial}
\end{eqnarray}
where $\phi$ and $\chi$ are 2-component objects.  We use the Pauli-Dirac
representation of the Dirac matrices
\begin{eqnarray}
 \gamma^i = \left( \begin{array}{cc}  0 & \sigma^i \\ -\sigma^i & 0
              \end{array} \right) \,, \qquad
\gamma^0 = \left( \begin{array}{cc}  1 & 0 \\ 0 & -1
              \end{array} \right)
\label{gamma}
\end{eqnarray}
where each block represents a $2\times2$ matrix, and $\sigma_i$s are
the Pauli matrices.  Note that the same $\gamma^\mu$  matrices are used which
we use in Cartesian co-ordinate system. For detail discussion about the Dirac
equation in cylindrical co-ordinates and the  form of the $\gamma^\mu$
matrices for cylindrical co-ordinate system see Ref. \cite{gre}. Multiplying
Eq. (\ref{dirac1}) by $\gamma^0$ from left, and using Eqs.
(\ref{trial}) and (\ref{gamma}) we get

\begin{eqnarray}
\left( \begin{array}{cc}E- M & \mathcal A \\ \mathcal A & E+M
\end{array} \right) \left( \begin{array}{c}
\phi(r) \\ \chi(r) \end{array} \right) =0\,, \label{d1}
\end{eqnarray}
where  $ \mathcal A= \boldsymbol{\sigma}.
\left( i\hat D_\mu - eQA_\mu  \right)=i \sigma^1 \partial_r +\frac{m -
eQ\alpha}{r}\sigma^2+\sigma^3p_z$.
Eq. (\ref{d1}) is divided into two coupled equations
\begin{eqnarray}
\phi (r) &=& \frac{\left[i \sigma^1 \partial_r +\frac{m -
eQ\alpha}{r}\sigma^2+\sigma^3p_z\right]}{E-M}  \chi (r)\,,
\label{eq1}\\\chi(r) &=& \frac{\left[i \sigma^1 \partial_r +\frac{m
- eQ\alpha}{r}\sigma^2+\sigma^3p_z\right]}{E+M} \phi(r) \,.
\label{eq2}
\end{eqnarray}
Eliminating $\chi(r)$, we obtain
\begin{eqnarray}
\phi(r) &=& \frac{\Big[i \sigma^1 \partial_r +\frac{m -
eQ\alpha}{r}\sigma^2+\sigma^3p_z \Big]^2}{E^2-M^2} \phi (r)\,.
\label{phieq1}
\end{eqnarray}
There will be two independent solutions for $\phi(r)$, which can be
taken, without any loss of generality, to be the eigen-states of
$\sigma_z$ with eigenvalues $s=\pm 1$. This means that we can choose
two independent solutions of the form
\begin{eqnarray}
\phi(r) = \left( \begin{array}{c} F_+(r) \\ 0 \end{array} \right)
\,, \qquad \phi(r) = \left( \begin{array}{c} 0 \\ F_-(r) \end{array}
\right) \,.
\label{phir1}
\end{eqnarray}
Since $\sigma^3\phi(r)= s\phi(r)$, Eq. (\ref{phieq1}) becomes
\begin{eqnarray}
\phi(r)\mid_{s=\pm 1} = \frac{-\frac{d^2}{dr^2} +\frac{(m -
eQ\alpha \mp 1/2)^2-1/4}{r^2} + p_z^2 }{E^2-M^2} \phi (r)
\label{p12}
\end{eqnarray}
For $s= +1$, using (\ref{phir1}) in (\ref{p12}), the differential
equation satisfied by $F_+$ is
\begin{eqnarray}
{d^2F_+ \over dr^2} + \left(\lambda^2 -\frac{\nu^2
-1/4}{r^2}\right)F_+ = 0 \,, \label{Fseqn2}
\end{eqnarray}
where
\begin{eqnarray}
\lambda = (E^2 - M^2 - p_z^2)^{1/2}, \nu = m - eQ\alpha - 1/2\,.
\label{usoln1}
\end{eqnarray}
%
It should be noted that, (\ref{Fseqn2}) can be considered as a well
known one dimensional Schr\"{o}dinger eigen-value equation with
inverse square interaction \cite{feher,tsutsui,veigy}.  It shows
classical scale symmetry and scaling anomaly, which will be
discussed separately in Sec. VI. Now we seek the solution of this
equation, which is of the form $ F_+ = r^{\frac{1}{2}}\mathscr
C_\nu(\lambda r)$. Where $\mathscr C$ denotes $J$, $Y$, $H^{1}$,
$H^{2}$ or any linear combination of these functions with constant
coefficients. From (\ref{eq2}), we find the two lower components and
write the spinor as
\begin{eqnarray}
U_+ (r, p_z) \equiv \left( \begin{array}{c}  r^{\frac{1}{2}}\mathscr
C_{\nu}(\lambda r) \\[2ex] 0 \\[2ex]  {\strut\textstyle p_z \over
\strut\textstyle E+ M} r^{\frac{1}{2}}\mathscr C_{\nu}(\lambda r) \\[2ex]
{\strut\textstyle i\lambda \over \strut\textstyle  E+ M}  r^{\frac{1}{2}}
\mathscr C_{\nu +1}(\lambda r)
\end{array} \right) \,,
\label{usoln4}
\end{eqnarray}
where the normalization has not been specified.  For $s = - 1$, we
get the same  Bessel differential equation (\ref{Fseqn2}), with
$\nu$ replaced by $\nu+1$. The solution, which can be obtained
similarly for this case  also, is of the  form
\begin{eqnarray}
U_- (r, p_z) \equiv \left( \begin{array}{c}  0 \\[2ex] r^{\frac{1}{2}}\mathscr
 C_{\nu +1}(\lambda r) \\[2ex]  {\strut\textstyle - i\lambda \over
 \strut\textstyle  E+ M}  r^{\frac{1}{2}} \mathscr C_{\nu }(\lambda r) \\[2ex]
 {\strut\textstyle - p_z \over \strut\textstyle E+ M} r^{\frac{1}{2}}\mathscr
 C_{\nu + 1}(\lambda r)
\end{array} \right) \,,
\label{usoln5}
\end{eqnarray}
A similar procedure can be adopted for negative frequency spinors.
In this case, it is easier to start with the two lower components
first and then find the upper components. The negative energy
spinors are found to be 
\begin{eqnarray}
V_+ (r, p_z) \equiv\left( \begin{array}{c}  {\strut\textstyle  p_z \over
\strut\textstyle E+ M} r^{\frac{1}{2}}\mathscr C_{\nu }(\lambda r) \\[2ex]
{\strut\textstyle i\lambda \over \strut\textstyle  E+ M} r^{\frac{1}{2}}
\mathscr C_{\nu +1}(\lambda r)  \\[2ex]  r^{\frac{1}{2}}\mathscr C_{\nu
}(\lambda r) \\[2ex] 0
\end{array} \right) \,,
\label{vsoln1}\\
V_- (r,p_z) \equiv \left( \begin{array}{c}  {\strut\textstyle -i\lambda \over
\strut\textstyle  E+ M} r^{\frac{1}{2}} \mathscr C_{\nu}(\lambda r) \\[2ex]
{\strut\textstyle - p_z \over \strut\textstyle E+ M} r^{\frac{1}{2}}\mathscr
C_{\nu +1}(\lambda r)  \\[2ex]  0 \\[2ex]r^{\frac{1}{2}}\mathscr C_{\nu
+1}(\lambda r)
\end{array} \right) \,.
\label{Vsoln2}
\end{eqnarray}

\section{SAE of radial Hamiltonian}

We now discuss the self-adjointness problem of our system. Therefore
we need to know the radial eigenvalue problem, which can be obtained
from  (\ref{dirac1}). It can be shown that the radial eigenvalue
equation is of the following form
\begin{eqnarray}
 H(r)S(r) = E S(r)\,,
\label{eigen1}
\end{eqnarray}
where the radial Hamiltonian and the eigenfunction are given by
\begin{eqnarray}
H(r) =\left( \begin{array}{cc} M & -\mathcal A \\ -\mathcal A & -M
\end{array} \right), S(r) =\left( \begin{array}{c} \phi(r) \\
\chi(r) \end{array} \right) \,, \label{eigen2}
\end{eqnarray}
where $ \mathcal A=  i \sigma^1 \partial_r +\frac{m -
eQ\alpha}{r}\sigma^2+\sigma^3p_z$. The differential operator $H(r)$
is symmetric over the domain
%
$D(H)=\psi(r)$, where $\psi(r)\in L_2(dr;C^4)$ and $\psi(0)=0$. This
means that for $\psi_1,\psi_2 \in D(H)$, the radial Hamiltonian
$H(r)$ satisfies the condition
 \begin{eqnarray}
\int_0^{\infty} dr \psi_1^{\dagger}(r)H(r)\psi_2(r) =\int_0^{\infty} dr
\left[H(r)\psi_1(r)\right]^{\dagger}\psi_2(r)\,.
\label{syme}
\end{eqnarray}
However, in domain $D(H)$  the radial Hamiltonian $H(r)$ is not
self-adjoint. A symmetric Hamiltonian is self-adjoint if its domain
coincides with that of the domain of its adjoint, i.e, $D(H)$ =
$D(H^\dagger)$.  The condition at the origin makes the Hamiltonian
non self-adjoint. The domain of the adjoint Hamiltonian
$H^{\dagger}(r)$ is
%
$D(H^{\dagger})= \psi(r)$, where $\psi(r)\in L_2(dr;C^4)$. We see
that $D(H)\neq D(H^\dagger)$, indicating that $H(r)$ is not
self-adjoint. To make the Hamiltonian self adjoint \cite{reed} we
use von Neumann's method of deficiency indices. It requires the
construction of eigen-space $D^{\pm}$ of $ H^{\dagger}$ with
eigenvalue $\pm i M~ ( M \neq 0$ is inserted for dimensional
reason). The up spinors (particle state) for the eigen-space $D^\pm$
are
\begin{eqnarray}
\phi^{\pm} = N\left( \begin{array}{c} r^{\frac{1}{2}}H^{1,2}_{\nu}(e^{\pm
i\frac{\pi}{2}}\lambda_1 r) \\[2ex] 0 \\[2ex]  {\strut\textstyle p_z \over
\strut\textstyle M(1\pm i)} r^{\frac{1}{2}}H^{1,2}_{\nu}(e^{\pm
i\frac{\pi}{2}}\lambda_1 r) \\[2ex] {\strut\textstyle ie^{\pm
i\frac{\pi}{2}}\lambda_1 \over \strut\textstyle M(1\pm i)} r^{\frac{1}{2}}
H^{1,2}_{\nu +1}(e^{\pm i\frac{\pi}{2}}\lambda_1 r)
\end{array} \right) \,,
\label{phi+-}
\end{eqnarray}
where $\lambda_1 = \left(2M^2 + p^2_z \right)^{\frac{1}{2}}$ and $N$
is the normalization constant. Here we have written $\phi^\pm$ for
up spinor (particle state) only. Similarly we will get $\phi^\pm$
for all other spinors. Since we will do our calculations for spin up
particle state only,  (\ref{phi+-}) is sufficient for us. Looking at
the asymptotic form of the Hankel functions
\begin{eqnarray}\label{assy}
\nonumber H^1_\nu(z)\to\left[\frac{2}{\pi z}\right]^{1/2}e^{i( z
-\frac{1}{2}\nu\pi -\frac{1}{4}\pi)},\\
~~~~~~~~~~~~~~~~~~~\mbox{for}~(-\pi<\arg z<2\pi)\\ \nonumber
H^2_\nu(z)\to\left[\frac{2}{\pi z}\right]^{1/2}e^{-i( z -\frac{1}{2}\nu\pi
-\frac{1}{4}\pi)},\\ ~~~~~~~~~~~~~~~~~~~\mbox{for}~(-2\pi<\arg z<\pi)
\label{assy1}
\end{eqnarray}
we find that the upper end of the integrals for evaluating the norms
of $\phi^\pm$ are finite for any $\nu$. However, near $r = 0$, the
Hankel function behavior can be found from the short distance
behavior of the Bessel function
\begin{eqnarray}
J_\nu(z) \to \frac{z^\nu}{2^\nu\Gamma(1+\nu)}, ~~~(\nu\neq -1, -2,
-3,...) \label{assy0}
\end{eqnarray}
Considering all components of the spinor of (\ref{phi+-}), we find
that $\phi^\pm$ are square integrable only in the interval
\begin{eqnarray}
 0 \leq \nu^2 < 1.
\label{def3}
\end{eqnarray}
Since beyond this range  there is no square integrable solutions
$\phi^\pm$, the deficiency indices, which are dimensions of the
eigen-space $D^\pm$,
\begin{eqnarray}
 n_\pm = dim (D^\pm),
\label{def4}
\end{eqnarray}
are zero , i.e,
\begin{eqnarray}
n_+ = n_- = 0.
\label{def5}
\end{eqnarray}
The closure of the operator $H(r)$ is the self-adjoint extension for
the case  (\ref{def5}).  We  therefore concentrate for the interval
(\ref{def3})  to carry out self-adjoint extensions. The existence of
complex eigenvalues for $H^{\dagger}(r)$ emphasizes the lack of
self-adjointness. The self-adjoint extensions of $H(r)$ are labeled
by the isometries $D^+\rightarrow D^-$, which can be parameterized
by
\begin{eqnarray}
\label{para}
\phi^+(r)\rightarrow e^{i\omega}\phi^-(r)
\end{eqnarray}
The correct domain ( It is better to call projection of the domain on spin up
particle state direction rather than only domain because we are concentrating
on spin up particle state only here ) for the self-adjoint extension
$H^{\omega}(r)$ of $H(r)$ is then given by
\begin{eqnarray}
D^{\omega}(H^\omega) \equiv D(H)+ \phi^+(r) +
e^{i\omega}\phi^-(r)\,, \label{self}
\end{eqnarray}
where $\omega\in \mathbb{R} (\bmod 2\pi)$. In the next section we
find out the bound state solutions using the domain
$D^{\omega}(H^\omega)$.

\section{solutions of radial Hamiltonian}

From the trial solution  (\ref{trial}) it is clear that the spinors
along  the $z$-direction is free,  as it should be, because there is
no constraint in the $z$-direction. We  therefore try to investigate
whether the spinors are bound in the radial direction. From now on
bound state means bound in the radial direction. Throughout our
calculation, we will use the  spin up state. Calculation for all
other spinor states are similar. From the general spin up state
(\ref{usoln4}) it is easy to see that it  serves as a square
integrable spin up state if we use $\mathscr C_\nu(\lambda r) =
H^1_\nu(\lambda r)$  and if $\lambda = (E^2 - M^2 -
p_z^2)^{\frac{1}{2}} = iq $, where $q$ is real positive. Similarly
we could have used $\mathscr C_\nu(\lambda r) = H^2_\nu(\lambda r)$
as square integrable function if we took $\lambda = (E^2 - M^2 -
p_z^2)^{\frac{1}{2}} = -iq$, where $q$ is real positive. We will use
Hankel function of the first kind $H^1_\nu$ to express bound state
solution. The bound state spinor of spin up particle is then found
to be
\begin{eqnarray}
U_+ (r, p_z) = B \sqrt{E+M} \left( \begin{array}{c}
r^{\frac{1}{2}}H^1_{\nu}(\lambda r) \\[2ex] 0 \\[2ex]  {\strut\textstyle p_z
\over \strut\textstyle E+ M} r^{\frac{1}{2}}H^1_{\nu}(\lambda r) \\[2ex]
{\strut\textstyle i\lambda \over \strut\textstyle  E+ M} r^{\frac{1}{2}}
H^1_{\nu +1}(\lambda r)
\end{array} \right) \,,
\label{ubound1}
\end{eqnarray}
where $B$ is the normalization constant. To find out bound state
eigenvalue we have to match the limiting value $r\rightarrow 0$ of
spinor  (\ref{ubound1}) with the limiting value $r\rightarrow 0$ of
the domain  (\ref{self}). For the sake of simplicity, we  set $p_z =
0$ before matching at the origin. The relevant range of $\nu$ is
given in  (\ref {def3}). In this range, the leading $r$-behavior of
the domain  (\ref{self})  for small $r$ is given by
\begin{eqnarray}
\psi(r) +\phi^+(r) + e^{i\omega}\phi^-(r) \rightarrow \left( \begin{array}{c}
A(\lambda_1)r^{\nu +1/2}  \\[2ex] 0 \\[2ex] 0 \\[2ex] D(\lambda_1)r^{-\nu -
1/2}
\end{array} \right) \,,
\label{blimit1}
\end{eqnarray}
where $A(\lambda_1) =
N\frac{i}{\sin\nu\pi}\frac{\lambda_1^\nu}{2^\nu\Gamma(1+\nu)}
\left[e^{-\frac{\pi\nu i}{ 2}} - e^{i\omega}e^{\frac{\pi\nu i}{ 2}}
\right]$, $D(\lambda_1) = -
Ni\frac{i}{\sin(\nu+1)\pi}\frac{\lambda_1^{-\nu -1}}{2^{-\nu
-1}\Gamma(-\nu)}\left[e^{-\frac{\pi\nu i}{ 2} -\frac{\pi i}{4}} -
e^{i\omega}e^{\frac{\pi\nu i}{ 2} + \frac{\pi i}{4}} \right]$,
and the leading term of the spinor  (\ref{ubound1}) is
\begin{eqnarray}
U_+(r, p_z) \rightarrow \left( \begin{array}{c} \tilde A(\lambda)r^{\nu +1/2}
\\[2ex] 0 \\[2ex] 0 \\[2ex]
\tilde D(\lambda)r^{-\nu  - 1/2}
\end{array} \right) \,,
\label{u_limit1}
\end{eqnarray}
where
$\tilde A(\lambda) =
B\sqrt{E+M}\frac{i}{\sin\nu\pi}\frac{\lambda^\nu}{2^\nu\Gamma(1+\nu)}
e^{- \pi\nu i}$, $\tilde D(\lambda) = - Bi\sqrt{E-
M}\frac{i}{\sin(\nu+1)\pi}\frac{\lambda^{-\nu -
1}}{2^{-\nu-1}\Gamma(-\nu)}$.
Since $U(r,p_z)\in D^\omega(H^\omega)$, the coefficients of leading
powers of $r$ in  (\ref{blimit1}) and   (\ref{u_limit1}) must match
and this gives the eigenvalue equation
\begin{eqnarray}
 \frac{\left( 1 + \frac{E}{M}\right)^{\nu +1}}{\left(1 -
 \frac{E}{M}\right)^{-\nu}} = - 2^{\nu +1/2}\frac{\sin\left(\omega/2 +
 \pi\nu/2\right)} {\sin\left(\omega/2 + \pi\nu/2 + \pi/4 \right)}
\label{cond1}
\end{eqnarray}
The left hand side of  (\ref{cond1}) is positive. So right hand side
should be positive and to ensure that  we impose  the condition
$\cot(\frac{\omega}{2} + \frac{\pi\nu}{2}) < -1 $, which is the
bound state condition.  Similarly we may get all other spinors for
the bound states. We move to the next section for the discussion of
scattering state (SS) solutions.

\section{SS solutions of radial Hamiltonian}

The scattering  state spinor of spin up particle is
\begin{eqnarray}
U_+ (r, p_z) = B\sqrt{E + M} \left( \begin{array}{c}  r^{\frac{1}{2}}A_{\nu}
\\[2ex] 0 \\[2ex]  {\strut\textstyle p_z \over \strut\textstyle E+ M}
r^{\frac{1}{2}}A_{\nu} \\[2ex] {\strut\textstyle i\lambda \over
\strut\textstyle  E+ M} r^{\frac{1}{2}}A_{\nu +1}
\end{array} \right) \,,
\label{uubound2}
\end{eqnarray}
where B is the normalization constant, $A_\nu = \left[a(\lambda)
J_{\nu}(\lambda r) + b(\lambda) J_{-\nu}(\lambda r)\right] $ and $
A_{\nu +1} = \left[\tilde a(\lambda) J_{\nu +1}(\lambda r) - \tilde
b(\lambda) J_{-\nu -1}(\lambda r)\right] $, $a(\lambda)$,
$b(\lambda)$, $\tilde a(\lambda)$ and $\tilde b(\lambda)$ are
constant coefficients.  To find out eigenvalue for the  scattering
state we have to match the limiting value $r\rightarrow 0$ of the
spinor  (\ref{uubound2}) with  (\ref{self}). For simplicity of
calculation we  set $p_z = 0$  in  (\ref{uubound2}) and (\ref{self})
before matching. In the limit $r\rightarrow 0$, the spinor
(\ref{uubound2}) looks  like (\ref{u_limit1}) but now the
coefficients of different powers of $r$ are,
$\tilde A(\lambda) =
B\sqrt{E+M}a(\lambda)\frac{\lambda^\nu}{2^\nu\Gamma(1+\nu)}$,
$\tilde D(\lambda) = - Bi\sqrt{E- M}\tilde
b(\lambda)\frac{\lambda^{-\nu - 1}}{2^{-\nu-1}\Gamma(-\nu)}$
and the limit $r\rightarrow 0$ of (\ref{self}) is given in
(\ref{blimit1}). Again equating the respective coefficients and
comparing between them we get the eigenvalue equation
\begin{eqnarray}
\nonumber \frac{\left( E + M\right)^{1/2}}{\left(E - M
 \right)^{1/2}}\frac{a(\lambda)}{\tilde
 b(\lambda)}\left(\frac{\lambda}{\lambda_1}\right)^{2\nu +1} = \\-
 \frac{\sin\left(\omega/2 + \pi\nu/2\right)} {\sin\left(\omega/2 + \pi\nu/2 +
 \pi/4\right)}
\label{cond2}
\end{eqnarray}
The right hand side of (\ref{cond2}) has to be positive in order to
get scattering state solutions. Similarly we may get all other
spinors for scattering states.  We now want to show the relation
between scattering state and bound state \cite{kumar,lan} for
completeness of our calculation. To calculate that we  expand
(\ref{uubound2}) in the limit $r\rightarrow \infty$. The leading
term in the asymptotic expansion of $U_+ (r, p_z)$, without
normalization  is given by
\begin{eqnarray}
U_+ (r, p_z) \rightarrow  \left( \begin{array}{c}  A(\lambda)e^{i\lambda r} +
B(\lambda)e^{-i\lambda r} \\[2ex] 0 \\[2ex]  {\strut\textstyle p_z \over
\strut\textstyle E+ M} A(\lambda)e^{i\lambda r} +{\strut\textstyle p_z \over
\strut\textstyle E+ M} B(\lambda)e^{-i\lambda r} \\[2ex] C(\lambda)e^{i\lambda
r} + D(\lambda)e^{-i\lambda r}
\end{array} \right) \,,
\label{uubound4}
\end{eqnarray}
where different coefficients are found to be $A(\lambda) =
\sqrt{\frac{E+M}{2\pi\lambda}}\left[a(\lambda)e^{-i\frac{\pi}{
2}(\nu+1/2)} + b(\lambda)e^{i\frac{\pi}{2}(\nu - 1/2)}\right]$,
$B(\lambda) =
\sqrt{\frac{E+M}{2\pi\lambda}}\left[a(\lambda)e^{i\frac{\pi
}{2}(\nu+1/2)} + b(\lambda)e^{-i\frac{\pi}{2}(\nu - 1/2)}\right]$,
$C(\lambda) =i\sqrt{\frac{E-M}{2\pi\lambda}}\left[\tilde
a(\lambda)e^{-i\frac{\pi }{2}(\nu+3/2)} - \tilde
b(\lambda)e^{i\frac{\pi}{2}(\nu + 1/2)}\right]$, $D(\lambda)
=i\sqrt{\frac{E-M}{2\pi\lambda}}\left[\tilde a(\lambda)e^{i\frac{\pi
}{2}(\nu+3/2)} - \tilde b(\lambda)e^{-i\frac{\pi}{2}(\nu +
1/2)}\right]$.
Now it is easy to see from the asymptotic expansion (\ref{uubound4})
that $e^{-i\lambda r}$ blows up on the positive imaginary
$\lambda$-axis but the other part $e^{i\lambda r}$ decays
exponentially there. So it reasonable to set the coefficients
$B(\lambda)$ and  $D(\lambda)$ zero for  purely positive imaginary
$\lambda $ to get square integrable behavior at $\infty$. This
corresponds to bound state. Using  (\ref{cond2}) and
(\ref{uubound4}) it can be shown that  the bound state eigenvalue is
again given by  (\ref{cond1}). Similarly it can also be shown that
for purely negative  imaginary $\lambda$ it is reasonable to set the
coefficients $A(\lambda)$ and  $C(\lambda)$ zero in order to get
bound state solution.

\section{Implication of scaling anomaly}

As pointed out in section II, we now discuss the classical scale
symmetry and  scaling anomaly of (\ref{Fseqn2}). To do that, we
rewrite (\ref{Fseqn2}) in the form of time independent
Schr\"{o}dinger equation
\begin{eqnarray}
\mathcal{H}_rF_s(r)=\mathcal{E}F_s(r)\,, \label{shamiltonian}
\end{eqnarray}
where the Hamiltonian $\mathcal{H}_r=-\partial^2_r -
(\nu^2-1/4)/r^2$, $\nu= m- eQ\alpha-1/2$ and the eigen-value
$\mathcal{E}=E^2-M^2$. Due to the cylindrical symmetry we just
consider the problem on $x$-$y$ plane, by setting $p_z=0$. This is a
well known one dimensional inverse square problem, discussed in many
areas of physics, from molecular physics to black hole
\cite{kumar,biru,kumar2,kumar3,bh,bh1,stjep,gov,karat,
giri,giri1,giri3,giri2}. Classically  (\ref{shamiltonian}) is scale
covariant. The scale transformation $r\to\xi r$ and $t\to\xi^2 t$
($\xi$ is scaling factor) transforms the the Hamiltonian
$\mathcal{H}_r\to (1/\xi^2)\mathcal{H}_r$. Scale covariance of
$\mathcal{H}_r$ implies that it should not have any bound state.
But, it is well known that there exists  a 1-parameter family
self-adjoint extensions (SAE) of $\mathcal{H}_r$ and due to this
SAE, a single bound state is formed. The  bound state energy for
$\nu^2\in[0,1)$ is given by
\begin{eqnarray}
\mathcal E=E^2-M^2= - \sqrt[\nu]{\frac{\sin(\Sigma/2+
3\pi\nu/4)}{\sin(\Sigma/2+\pi\nu/2)}}\,,\label{seigenvalue}
\end{eqnarray}
where $\Sigma$ is the self-adjoint extension parameter for
$\mathcal{H}_r$. Existence of this bound state immediately breaks
the scale symmetry, which leads to scaling anomaly.

If we assume that the energy of the Dirac particle $E$ to be real,
then from the bound state condition $\mathcal E<0$, we get a bound
on the energy of the Dirac particle to be $E\in(-M,M)$. This bound
still holds for the free Dirac particle, which we get by taking
limit $\alpha\to 0$ in (\ref{seigenvalue}).

Quantum mechanically, scale transformation is associated with a
generator, called scaling operator $\Lambda = \frac{1}{2} (rp_r +
p_r r)$, where $p_r= -i\frac{d}{dr}$. It can be shown that the
action of the operator $\Lambda$ on a generic element of the self
adjoint domain $D_\Sigma(\mathcal {H}_r)$ does throw the element
outside the domain for some values of the self-adjoint extension
parameter $\Sigma$. This indicates that there is  scaling anomaly,
occurred due to self-adjoint extensions. However, it can be shown
that for $\Sigma = -\nu \pi/2$ and $ \Sigma = -3\nu \pi/2$, the
action of the operator $\Lambda$ on any element of the domain
$D_\Sigma(\mathcal{H}_r)$ does not throw it outside the domain. So,
in these two cases scaling symmetry is still restored even after
self-adjoint extensions and thus the bound state does not occur.

\section{Conclusion and discussion}

In this paper we  calculated the solution of the Dirac equation in
the field of an infinitely long solenoid. We  showed that there is
nontrivial bound state and as well as scattering state solution in
the range $\nu^2\in[0,1)$. We  showed only spin up particle state
details of self-adjoint extensions in our calculation, but for all
other spinors calculations are similar.  We point out that for $\nu
= 0$, the solution of $H^\dagger\phi^{\pm} = \pm iM\phi^{\pm}$
involves $r^{\frac{1}{2}}H^1_1(i\lambda_1r)$, which is not square
integrable at the origin. So at $\nu = 0$ deficiency indices are
zero, i.e, $n_+ = n_- = 0$, that means closure of domain is the
self-adjoint extension for $\nu = 0$. We have projected the whole
problem from 4-component Hilbert space $(L_2(dr d\phi dz; \mathcal
C^4))$ to 1-component Hilbert space $(L_2(dr d\phi dz; \mathcal C))$
keeping in mind that we  make the whole $(4\times 4)$ Dirac
Hamiltonian self-adjoint unlike Ref.~\cite{gitman}, where they
discussed the self-adjoint issue projecting the problem on $L_2(r dr
d\phi dz; \mathcal C^2)$.  The significant difference between these
two approaches is that in our case we  have  a 1-parameter family of
self-adjoint extensions unlike Ref.~\cite{gitman} where they have
2-parameter family of self-adjoint extensions.  As in
Ref.~\cite{kumar,lan}  we  studied that there is a relation between
scattering state and bound state and it occurs for the purely
imaginary value of $\lambda$. This has been done by demanding square
integrability of the scattering state at spatial infinity. We showed
that scattering state eigenvalue reduces to bound state eigenvalue
for purely imaginary  $\lambda$. Finally we discussed the
implications of scaling anomaly on Dirac particle in background
solenoid field. The energy of the Dirac particle in solenoid field
background has bounds from both sides $E\in(-M,M)$. By taking
appropriate limit, these bounds were also shown to hold for free
Dirac particles.

\section{Acknowledgment}

We thank Kumar S. Gupta and Palash B. Pal for comments on the manuscript and
helpful discussions.

\end{document}